# New Perspectives for Rashba Spin-Orbit Coupling

A. Manchon[1], H.C. Koo[2,3], J. Nitta[4], S.M. Frolov[5], R.A. Duine[6]

[1]King Abdullah University of Science and Technology (KAUST), Physical Science and Engineering Division, Thuwal 23955-6900, Saudi Arabia
[2]Center for Spintronics Research, Korea Institute of Science and Technology (KIST), 39-1 Hawolgok-dong, Seongbukgu, Seoul, 136-791, Korea
[3]KU-KIST Graduate School of Converging Science and Technology, Korea University, Seoul, 136-701, Korea
[4]Department of Materials Science, Tohoku University, 980-8579 Sendai, Miyagi, Japan
[5]Department of Physics and Astronomy, University of Pittsburgh, Pittsburgh, PA, 15260, USA
[6]Institute for Theoretical Physics and Center for Extreme Matter and Emergent Phenomena, Utrecht University, Leuvenlaan 4, 3584 CE Utrecht, The Netherlands

**In 1984, Bychkov and Rashba introduced a simple form of spin-orbit coupling to explain certain peculiarities in the electron spin resonance of two-dimensional semiconductors. Over the past thirty years, similar ideas have been leading to a vast number of predictions, discoveries, and innovative concepts far beyond semiconductors. The past decade has been particularly creative with the realizations of means to manipulate spin orientation by moving electrons in space, controlling electron trajectories using spin as a steering wheel, and with the discovery of new topological classes of materials. These developments reinvigorated the interest of physicists and materials scientists in the development of inversion asymmetric structures ranging from layered graphene-like materials to cold atoms. This review presents the most remarkable recent and ongoing realizations of Rashba physics in condensed matter and beyond.**

# Introduction

In crystals lacking an inversion center, electronic energy bands are split by spin-orbit (SO) coupling. The Rashba SO coupling, a SO coupling linear in momentum $p$, was originally proposed for noncentrosymmetric wurtzite semiconductors [1]. After the establishment of modulation-doped semiconductor hetrostructures, Bychkov and Rashba applied it to the SO coupling in a two-dimensional electron gas (2DEG) with structural inversion asymmetry [2]. In systems with inversion symmetry breaking, SO coupling becomes *odd* in momentum $p$ which, in the simplest two-dimensional free electron approximation, reduces to a linear dependence [2]. Odd-in-$p$ SO coupling has been confirmed in a wide variety of materials lacking spatial inversion (see Box 1). The essential feature of any SO coupling is that electrons moving in an electric field experience, even in the absence of an external magnetic field, an effective magnetic field in their frame of motion, called the SO field, which couples to the electron's magnetic moment. In the case of a system with inversion symmetry breaking this SO field becomes *odd* in the electron's momentum $p$, which enables a wide variety of fascinating phenomena (see Box 2). By extension, in this review we refer to this odd-in-$p$ SO coupling as Rashba SO coupling. The exploration of Rashba physics is now at the heart of the growing research field of *spin-orbitronics*, a branch of spintronics [3] focusing on the manipulation of non-equilibrium materials properties using SO coupling (see Fig. 1). Here we review the most recent developments involving such (odd-in-$p$) Rashba SO interactions in various fields of physics and materials science.

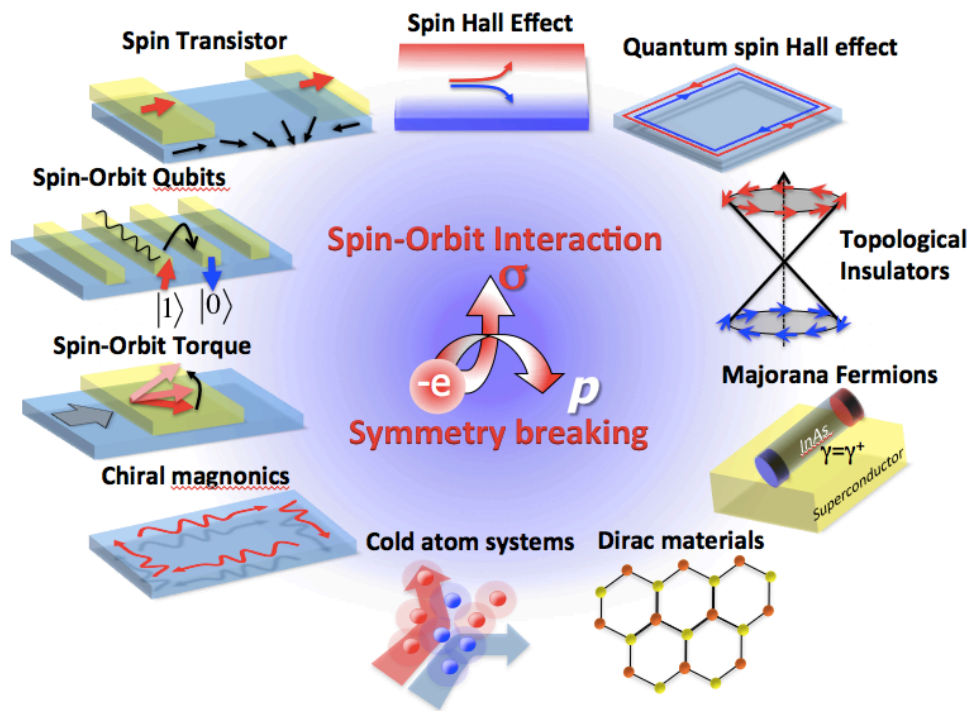

**Figure 1| Diagram showing various realizations of spin-orbitronics:** When SO coupling is present in systems with broken inversion symmetry, unique transport properties emerge giving birth to the tremendously active field of spin-orbitronics, the art of manipulating spin using SO coupling.

## Box 1| Rashba spin-orbit interaction
### Origin of Rashba Spin-Orbit Coupling

When an electron with momentum $\vec{p}$ moves across a magnetic field $\vec{B}$, it experiences a Lorentz force in the direction perpendicular to its motion $\vec{F} = -e\vec{p} \times \vec{B}/m$ and possesses Zeeman energy $\mu_B \vec{\sigma} \cdot \vec{B}$, where $\vec{\sigma}$ is the vector of Pauli spin matrices, $m$ and $e$ are the mass and charge of the electron, and $\mu_B = 9.27 \times 10^{-24}$ J/T is the Bohr magnetron. By analogy, when this electron moves across an electric field $\vec{E}$, it experiences an effective magnetic field $\vec{B}_{eff} \sim \vec{E} \times \vec{p}/mc^2$ in its rest-frame ($c$ is the speed of light), a field that also induces a momentum-dependent Zeeman energy $\hat{H}_{so} \sim \mu_B (\vec{E} \times \vec{p}) \cdot \vec{\sigma}/mc^2$, called the SO coupling. In crystals, the electric field is given by the gradient of the crystal potential, $\vec{E} = -\vec{\nabla}V$.

In quantum wells with structural inversion symmetry broken along the growth direction $\vec{z}$, the spin subbands are split in energy (see Fig. 2(a,b)). Such band splitting is also observed at certain metallic surfaces (see Fig. 2(d)) and was explained by Bychkov and Rashba considering an electric field $\vec{E} = E_z \vec{z}$ resulting in an effective SO coupling of the form [2]

$$\hat{H}_R = \frac{\alpha_R}{\hbar}(\vec{z} \times \vec{p}) \cdot \vec{\sigma}, \tag{B.1}$$

where $\alpha_R$ is called the Rashba parameter. Nevertheless this convenient form, derived for two-dimensional plane waves, is only phenomenological and does not apply *per se* on realistic systems. Indeed, theoretical investigations showed that the lack of inversion symmetry does not only create an additional electric field $E_z$ but also distorts the electron wave function close to the nuclei where the plane wave approximation is not valid [4]. In other words, in the solid state the Dirac gap $mc^2 \approx 0.5$ MeV is replaced by the energy gap $\approx 1$ eV between electrons and holes and $\alpha_R/\hbar >> \mu_B E_z/mc^2$. In addition, the inversion symmetry breaking only imposes the SO coupling to be *odd* in electron momentum $\vec{p}$, i.e. $\hat{H}_{SO} = \vec{w}(\vec{p}) \cdot \hat{\sigma}$ where $\vec{w}(-\vec{p}) = -\vec{w}(\vec{p})$, like in the case of *p*-cubic Dresselhaus SO coupling in zinc-blende III-V compound semiconductors [5]. It becomes linear in $\vec{p}$ only under certain conditions (e.g. when the free electron approximation is valid or under strain). Therefore in the discussion provided in the present review, one has to keep in mind that the *p*-linear Rashba SO coupling is a useful phenomenological approximation that does not entirely reflect the actual form of the SO coupling in inversion asymmetric systems.

### Measuring Rashba Spin-Orbit Coupling

The magnitude of the phenomenological Rashba parameter $\alpha_R$ has been estimated in a wide range of materials presenting either interfacial or bulk inversion symmetry breaking. The analysis of Shubnikov-de Haas oscillations and spin precession in InAlAs/InGaAs [6], [7] (see Fig. 2(c)) yields a Rashba parameter (~0.67 x $10^{-11}$ eV.m) comparable to recent estimations in LaAlO$_3$/SrTiO$_3$ heterointerfaces using weak localization measurements (~0.5 x $10^{-11}$ eV.m) [8], [9]. Signatures of Rashba SO coupling have also been confirmed at the surface of heavy metals such as Au [10] or Bi/Ag alloys [11], using angle-resolved photoemission spectroscopy (ARPES), and revealing a gigantic Rashba effect, about two orders of magnitude larger than in semiconductors (~3.7 x $10^{-10}$ eV.m for Bi/Ag alloy). More recently, topological insulators have been shown to display comparable Rashba parameters (~4 x $10^{-10}$ eV.m for Bi$_2$Se$_3$ [12]). Structures presenting bulk inversion symmetry breaking also show evidence of a Rashba-type SO splitting of the band structure. For instance, the polar semiconductor BiTeI displays a bulk Rashba parameter (~3.85 x $10^{-10}$ eV.m [13]) as large as on the surface of topological insulators.

## Spin generation, manipulation and detection

Charge carriers in materials with Rashba SO coupling experience a momentum-dependent effective magnetic field, a spin-dependent correction to velocity as well as a geometric phase resulting from the SO coupling (see Box 2). These features are particularly attractive for the realization of device concepts in which spin polarization is generated out of charge current, manipulated by electric fields and detected as voltage or Kerr rotation.

### *a. Spin Hall effect*

The spin Hall effect is the conversion of an unpolarized charge current into a chargeless pure spin current, i.e. a net spin flow without charge flow, transverse to it. This happens through two classes of mechanisms. In the first class, electrons with different spin projections diffuse towards opposite directions upon scattering against SO-coupled impurities. This spin-dependent extrinsic Mott scattering is at the core of the original prediction of spin Hall effect formulated forty years ago by D'yakonov and Perel [14], and more recently revived by Hirsch [15]. The second class concerns the spin-dependent distortion of the electrons trajectory in the presence of SO coupled band structure (see Box 2). This so-called intrinsic spin Hall effect has been recently put forward independently by Murakami et al. [16] and Sinova et al. [17]. In the latter work, a universal spin Hall conductivity $\sigma_H = e\Phi_B/8\pi^2$ was predicted in the case of a ballistic 2DEG with Rashba interaction, where $\Phi_B$ is the geometrical phase (also called Berry phase) which is acquired by a state upon being transported around a loop in momentum space [18].

The experimental observation of the spin Hall effect in bulk GaAs and strained InGaAs was demonstrated using Kerr rotation microscopy [19]. Without applying any external magnetic fields, an out-of-plane spin polarization with opposite sign on opposite edges of the sample was detected. The amplitude of the spin polarization was weak, and the mechanism was attributed to the extrinsic spin Hall effect. The spin Hall effect in a GaAs 2D hole system was observed by Wunderlich *et al*. [20] in light emitting diodes. The magnitude of the spin polarization is in agreement with the prediction of the intrinsic spin Hall effect. Spin transistors and spin Hall effects have been combined by realizing an all-semiconductor spin Hall effect transistor [21]. A spin AND logic function was demonstrated in a semiconductor channel with two gates. Here, spin polarized carriers are detected by the inverse spin Hall effect and the spin generation in this device was achieved optically. The search for large spin Hall effect has been extended to metals by Valenzuela and Tinkham [22] and is now one of the most active areas of spintronics. Since these SO effects do not necessitate inversion symmetry breaking and are hence not directly related to Rashba physics, we refer the reader to the excellent reviews available on this topic [23], [24], [25].

*b. Spin Interferences*

Spin-polarized electrons experiencing Rashba SO coupling acquire a geometrical (also called Berry) phase that may result in spin interferences (see Box 2). Indeed, The rotation operator for spin 1/2 produces a minus sign under $2\pi$ rotation [26]. Neutron spin interference experiments have verified this extraordinary prediction of quantum mechanics [27], [28]. A local magnetic field causes precession of the electron's spin in a way that depends on the path of the electron. Spin interference effects controlled by an electric field, distinct from conventional spin interference, have been demonstrated in a single HgTe ring [29] and in small arrays of mesoscopic InGaAs 2DEG rings [30] with strong Rashba SO coupling. This interference is an Aharonov-Casher [31] effect and is the electromagnetic dual to the Aharonov-Bohm [32] effect. The spin precession rate can be controlled in a precise and predictable way with an electrostatic gate [7].

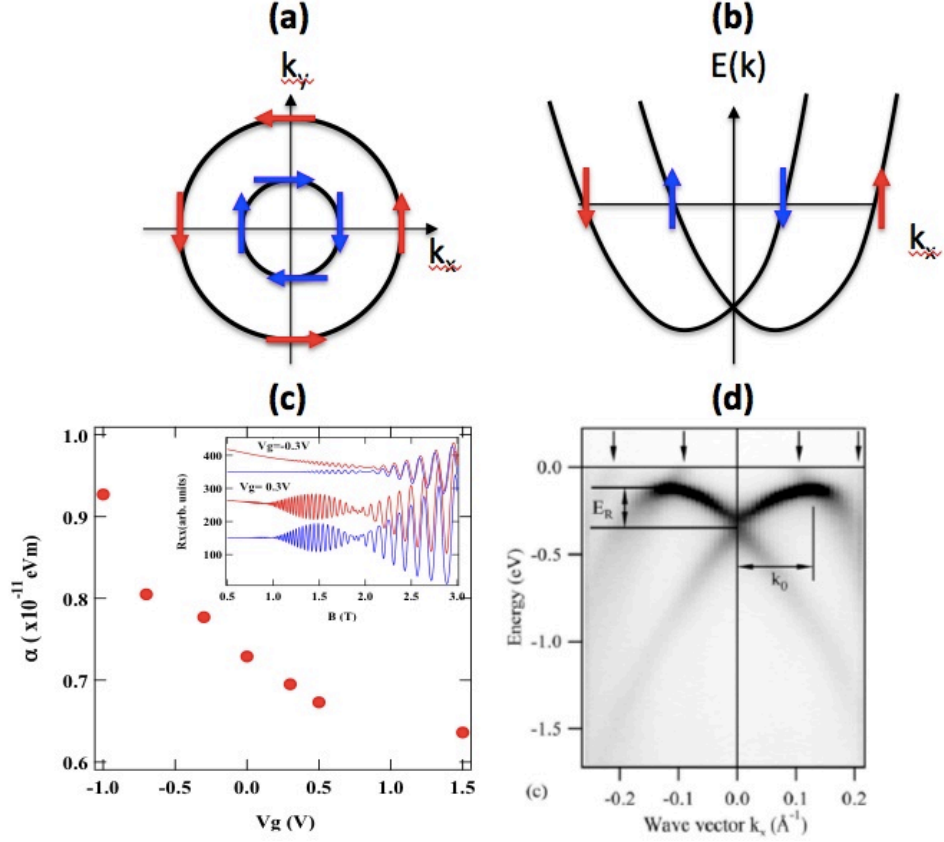

**Figure 2| Rashba spin splitting at interfaces** : (a) Schematics of the Fermi surface of a two-dimensional free electron gas with Rashba spin-orbit coupling: the spin angular momentum is locked on the linear momentum; (b) Schematics of the energy dispersion with spin-momentum locking; (c) Gate control of the magnitude of Rashba spin-orbit coupling parameter in InAlAs/InGaAs quantum well (from [7]) – Inset: Shubnikov-de Haas oscillations from which Rashba parameter is extracted; (d) Energy dispersion at the surface of Bi/Ag alloy measured by ARPES displaying a clear Rashba splitting (from [11]).

*c. Spin Galvanic effect*

The locking between the electron momentum and its spin angular momentum results in the so-called spin galvanic effect (see Box 2). Following the Rashba Hamiltonian Eq. (B.1), the spin galvanic effect is given by $\vec{j}_c = -e\alpha_R (\vec{z} \times \vec{S})/\hbar$, where $\vec{S}$ is the non-equilibrium spin density (created either electrically or optically) and $\vec{j}_c$ is the induced charge current density [33]. This concept was originally developed in the context of optical manipulation of spin in semiconductors and observed in quantum wells [34], [35]. The spin galvanic effect has also been recently realized in a NiFe/Ag/(Bi/Ag) lateral device [36]. In this system, a spin current is pumped from NiFe into Ag and converted into a transverse charge current through the spin galvanic effect that takes place at the Ag/(Bi/Ag) interface. The Onsager reciprocal of spin galvanic effect, the inverse spin galvanic effect (sometimes called the Edelstein effect [37]) has been recently observed in strained semiconductors [38] and quantum wells [39]. Following the Rashba symmetry the spin density generated by the current is $\vec{S} = \alpha_R m(\vec{z} \times \vec{j}_c)/e\hbar$. It is of particular significance for manipulating the magnetization of single ferromagnets (see below).

*d. Electrical spin manipulation*

An essential aspect that renders Rashba SO coupling particularly attractive for spintronics and quantum computation is its ability to be controlled by an external gate voltage placed on top of the 2DEG. Indeed, since the strength of the Rashba parameter is directly related to the interfacial potential drop (see Box 1), applying a gate voltage modifies the electron occupation, which in turn controls the magnitude of the Rashba SO coupling, as experimentally demonstrated in InGaAs/InAlAs heterostructures [7], [40]. The electric control of spin states is superior to the magnetic field control due to a better scalability, lower power consumption and the possibility for local manipulation of the spin states. The first spintronic device concept utilizing Rashba SO coupling was a spin field-effect transistor proposed by Datta and Das [41]. The implementation of this transistor relies on spin injection from a ferromagnetic electrode into a 2DEG and, subsequently, on gate controlled precession angle of the injected electron's spin. The experiment that combines spin injection and precession towards the spin field-effect transistor was performed by Koo et. al. [42].

A Stern-Gerlach spin filter was proposed by using the spatial gradient of the Rashba interaction [43]. A spatial gradient of the effective magnetic field due to the Rashba SO coupling causes a Stern-Gerlach type spin separation. Almost 100% spin polarization can be realized, even without applying any external magnetic fields or using ferromagnetic contacts. In contrast to the spin Hall effect, the spin-polarized orientation is not out-of-plane but in-plane. This inhomogeneous SO-induced electronic spin separation has been demonstrated in semiconductor quantum point contacts [44]. Such a spin-filter device can be used for electrical spin detection [45].

Electron spin resonance (ESR) using static and oscillating magnetic fields is utilized for manipulation of individual electron spins in quantum information processing [46]. The oscillating field induced by Rashba and Dresselhaus SO coupling is driven by the free motion of electrons that bounce at frequencies of tens of GigaHertz in μm-scale channels. Coherent control of individual electron spins using gigahertz electric fields by means of electric dipole spin resonance (EDSR) has been performed in GaAs/AlGaAs gate-defined quantum dots [47], and in InAs nanowires [48] establishing SO qubits. Rabi frequencies exceeding 100 MHz were demonstrated in InSb nanowires [49]. The demonstration of SO qubits coupled to superconducting resonators paves way for a scalable quantum computing architecture [50].

*e. Suppressing spin relaxation*

In the above discussion, we have highlighted the efficient coupling of the electron's spin to its motion enabled by Rashba interaction and the ways this can be used for spin control. On the downside, however, the momentum-changing scattering of an electron moving through a semiconductor causes sudden changes in the effective Rashba magnetic field leading to spin randomization [51]. Hence, suppressing spin relaxation in the presence of strong, tunable SO coupling is a major challenge of semiconductor spintronics.

In III-V semiconductor heterostructures, the Dresselhaus SO due to bulk inversion asymmetry also gives rise to a band spin splitting, given by contributions linear and cubic in momentum $p$. The most effective way to suppress the spin relaxation is to utilize the so-called persistent spin helix condition [52], [53] where the Rashba SO strength is equal to $p$-linear Dresselhaus SO strength. Under this condition, the spin polarization is preserved even after scattering events. This conservation is predicted to be robust against all forms of spin-independent scattering, including electron-electron interaction, but is broken by spin-dependent scattering and $p$-cubic Dresselhaus terms. The persistent spin helix in semiconductor quantum wells was confirmed by optical transient spin-grating spectroscopy by Koralek *et al.* [54]. These authors found that the spin lifetime is enhanced by two orders of magnitude near the exact persistent spin helix point. Recently, gate controlled spin helix states have been realized using a direct determination of the Rashba and Dresselhaus interactions ratio [55] and spin transistor design based on gate-tunable spin helix has been proposed [52].

**Box 2| Physics of the Rashba effect**

To discuss the physics induced by SO coupling in systems lacking inversion symmetry, let us consider the $p$-linear Rashba SO coupling, Eq. (B.1), introduced previously [2]. This equation describes a Zeeman term involving a magnetic field proportional to the electron momentum $p$. Consequently, when electrons flow along the $x$-axis, they experience an effective magnetic field along the $y$-axis, $B_{Ry}$, called the Rashba field, as depicted in Fig. B1 (top). The magnitude of the Rashba field can be calculated from $B_{Ry} = 2\alpha_R k_F / g\mu_B$, where $k_F$ and $g$ are the Fermi wavevector and $g$-factor of the carriers in the conduction channel, respectively.

**Rashba field and spin precession**

When the electron spin is not aligned with the Rashba field, spin precession takes place with a frequency that depends on the magnitude of the field. In Fig. B1 (bottom), the spin-polarized electrons injected along the $x$-axis precess under the influence of the Rashba field even without an applied magnetic field. The magnitude of the electric field, and hence the strength of the Rashba field and spin precession rate can be controlled by a gate voltage [7], [41], [42] (see Fig. 2(c)). In the diffusive regime, this precession is at the origin of the so-called D'yakonov-Perel spin relaxation mechanism [51]. An interesting consequence of the emergence of the Rashba field is the possibility to polarize flowing electron along the direction of this field. This effect, called *inverse spin galvanic effect* [37], has a counterpart referred to as *spin galvanic effect* [34], i.e. the conversion of non-equilibrium spin density (created by either optical or electrical means) into a charge current.

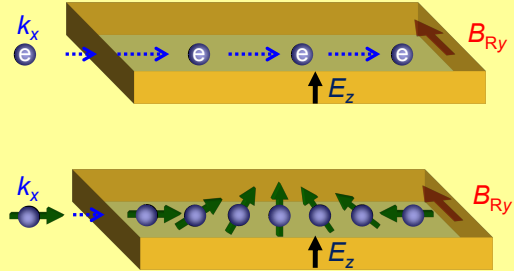

**Figure B1| Rashba spin-orbit interaction.** The moving electrons ($k_x$) with a perpendicular electric field ($E_z$) induce the Rashba field, $B_{Ry}$ (top). In a Rashba system, the spin-polarized electrons precess around the axis of the Rashba field (bottom).

**Berry curvature and spin Hall effect**

Rashba effect modifies the velocity according to $\vec{v}_a = \partial_{\vec{p}} \hat{H}_R = -(\alpha_R / \hbar)\vec{z} \times \hat{\sigma}$. Physically, the electrons trajectory is "bent" due to SO coupling, in a manner very similar to how Magnus force distorts the trajectory of spinning balls in classical mechanics. The direction of the distortion depends on the direction of the angular momentum (i.e. of the spinning), which results in a spin Hall effect (see Fig. 1). This additional velocity can be formulated in terms of an effective Lorentz force acting on the electron semiclassical wavepacket. The effective magnetic field that produces this Lorentz force is called Berry curvature [56] $\vec{\Omega}(\vec{p}) \propto \alpha_R \vec{\nabla}_p \times (\hat{\sigma} \times \vec{p})$ and only depends on the geometry of the band structure. The resulting anomalous velocity induces an off-diagonal conductivity that can be nonzero if time-reversal symmetry is broken. The language of Berry curvature and its associated Berry phase has been extremely successful in describing the various properties of Rashba and Dirac materials [56].

## Spin-orbit torque in ferromagnets

The electrical control of the magnetization direction of small magnets is currently among the most active areas in spintronics due to its interest for memory, logic and data storage applications [57]. For the past fifteen years, this control has been achieved through the transfer of spin angular momentum between a flowing spin current and the local magnetization of a ferromagnet [58], [59], as now conventionally observed in magnetic spin-valves structures and magnetic domain walls [60]. Less than ten years ago, an alternative mechanism based on the inverse spin galvanic effect has been proposed, which allows for a direct transfer between the carrier momentum to the local magnetization mediated by SO coupling.

Indeed, in non-centrosymmetric (e.g. wurtzite) magnetic semiconductors or asymmetrically grown ultrathin magnetic layers (e.g. a ferromagnet deposited on a heavy metal), the spin density generated by the inverse spin galvanic effect exerts a torque on the magnetization [61], [62], [63]. This so-called *SO torque* can be used to excite or reverse the magnetization direction and is therefore potentially useful for applications such as magnetic memories or logic [64]. This torque was first observed in (Ga,Mn)As [65] where straining the zinc-blende structure is responsible for the emergence of *p*-linear Dresselhaus SO coupling. Indications of the emergence of intrinsic (Berry phase-induced) SO torque has been reported recently [66]. A year later, SO torques were reported in Pt/Co/Alox asymmetric structures and attributed to interfacial Rashba SO coupling [67]. Since then, a wealth of experimental investigations have revealed the complex nature of the SO torque in magnetic semiconductors [66] and metallic multilayers [68], [69], [70]. A major difficulty is to clearly identify the physical origin of the SO torque, as spin Hall effect plays an important role in magnetic multilayers [71]. Several mechanisms, going beyond the inverse spin galvanic effect paradigm, have been proposed to explain the experimental results [66], [72], [73].

Recent material developments have allowed for the observation of very large SO torques at the interface with topological insulators [74], [75] (see below). Although the Dirac cone expected to emerge at the surface of these materials is probably dramatically altered by the presence of the ferromagnet, this result demonstrates the suitability of topological insulators in controlling SO torques. Another class of systems that might benefit from the emergence of SO torques are the antiferromagnets. It has been theoretically demonstrated that SO torque could be utilized to manipulate coherently the order parameter of these materials, which opens promising perspectives in the field of antiferromagnetic spintronics [76].

Finally, it is worth to mention that interfacial or bulk inversion symmetry breaking also has a dramatic impact on the transport properties of spin waves, resulting in coupling effects very similar to the electrons undergoing Rashba SO coupling. In such materials, the magnetic energy acquires an antisymmetric exchange interaction, known as Dzyaloshinskii-Moriya interaction [77], [78]. This interaction acts like a Rashba SO coupling on spin waves: the magnetization is distorted resulting in chiral magnetism and possibly skyrmions [79], but even more importantly for the present review, the magnon energy dispersion acquires a component linear in the magnon momentum, as observed experimentally [80]. This linear component in the dispersion relation induces an anomalous velocity leading to the magnon Hall effect [81], orbital moment and edge currents [82], as well as to chiral damping [83]. It was recently proposed that this antisymmetric exchange interaction enables a so-called magnon-driven torque displaying striking similarities with the electron-driven SO torque discussed above [84].

## Topological states and Majorana fermions

Spin-orbit interaction plays a central role as a design element of topological states of matter, both recently discovered and proposed. Here we review the topological insulator and topological superconductor states, both remarkable for their edge states, which are characterized by helical spin textures and Majorana fermions [85].

Topological insulators come in 2D and 3D varieties, with the 2D topological insulators, known also as quantum spin Hall insulators (Fig. 3(a)), discovered first in HgTe/CdTe quantum wells [86]–[88] (Fig. 3(b)). They were later also reported in InAs/GaSb quantum wells [89]. These compounds are composed of heavy elements and therefore exhibit strong bulk SO interaction. In the quantum spin Hall insulator, the bulk is insulating, while two one-dimensional conduction channels exist on each edge. On one edge, two channels are counter propagating while carrying opposite spin representing a helical spin pair (Fig. 3(a)). SO interaction is so strong in quantum spin Hall insulators that the top of the uppermost valence subband split from the other subbands by SO interaction is above the bottom of the lowest conduction band subband. In this inverted band structure a gap opens due to the interaction of the valence and conduction subbands, while SO interaction ensures an odd number of helical pairs on each edge of the system. As opposed to the quantum Hall insulator, the quantum spin Hall insulator exists at zero magnetic field, and the robustness of the edge modes is protected by time-reversal symmetry, a property preserved by SO interactions. Backscattering of the quantum spin Hall edge is strongly suppressed because scattering into a state of opposite momentum requires jumping to the opposite edge if spin orientation is preserved. Magnetic scattering is among factors that limit the coherence of quantum spin Hall edge states. In today's experiments these edge states are observed on the micron scale, as opposed to a millimeter scale for the quantum Hall edge states.

Three-dimensional topological insulators are an extension of the 2D concept to 3D. Again, heavy element compounds such as $Bi_2Se_3$, $Bi_xSb_{1-x}$, $Bi_2Te_3$, with strong bulk SO interaction, exhibit the topological insulator phase [85]. Instead of one-dimensional channels the edges carry a surface state characterized by a Rashba spin texture. Namely, spin is locked to momentum and always points perpendicular to it (see Eq. (B.1)). This spin texture was directly observed in ARPES experiments [12] (Fig. 3(c,d)). Spin-momentum locking also suppresses backscattering, which has been reported in scanning tunneling microscopy experiments [90].

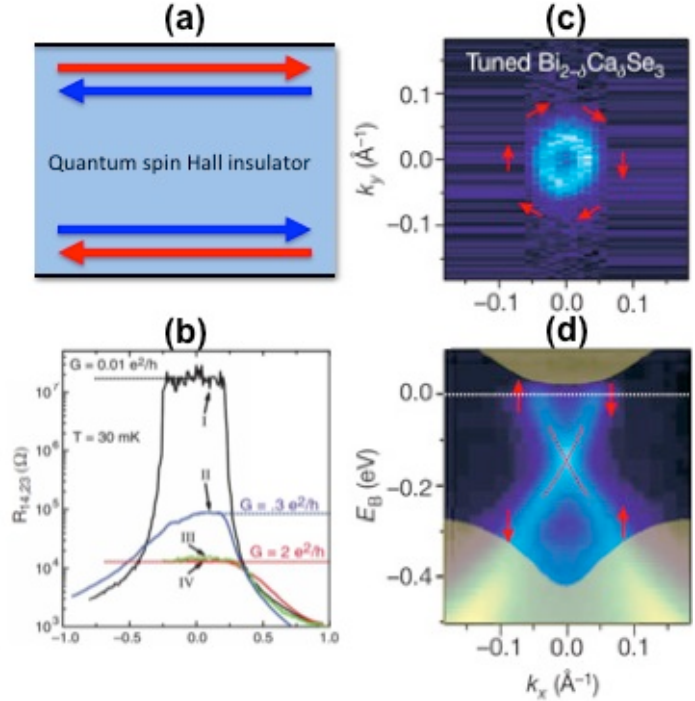

**Figure 3| Topological insulators:** (a) Schematics of the quantum spin Hall insulator, which is insulating in the bulk and supports pure spin current flow at its edges. (b) Experimental evidence of the quantum spin Hall effect in HgTe quantum wells [87]. (c) Two-dimensional mapping of the Fermi surface and (d) band structure of a topological insulator measured by ARPES [91].

Topological superconductivity can be understood by a simple formula: topological insulator plus superconductivity [92]. Topological superconductivity can in principle be intrinsic to a compound, or induced by proximity to a non-topological superconductor. Topological superconductors are characterized by an inverted superconducting gap, though at the moment it is not clear how to detect the sign of the gap experimentally. The most remarkable manifestations of topological superconductivity are related to its edge states and derive from the properties of the edge states of a topological insulator. Superconductivity transforms electrons in the edge states into Bogoliubov quasiparticles, which necessarily possess particle-hole symmetry. Namely, if there is a state at positive energy there must be a state at negative energy of the same magnitude, with zero energy being the Fermi level. If there are an odd number of states, particle-hole symmetry dictates that one of the states must be pinned to zero energy. This is the case in topological superconductors due to the odd number of helical pairs on each edges of the system. The zero state or mode is then known as a Majorana fermion, because it corresponds to its own antiparticle [93]. Realizing Majorana fermions by combining 2D and 3D topological insulators with conventional superconductors is an active research area [94].

Initial experimental evidence of Majorana fermions was, however, obtained in one-dimensional systems, i.e. not starting from 2D or 3D topological insulating phases. Nevertheless, SO interaction has been the key ingredient in this case as well. A one-dimensional wire of InSb, a semiconductor with strong Rashba SO interaction, has an electronic spectrum that consists of two spin-resolved parabolas shifted in opposite directions in momentum space (Fig. 4(a)). Applying an external magnetic field perpendicular to the intrinsic Rashba field mixes the two subbands and opens a gap at the crossing point. If the Fermi level is inside this gap, we obtain a helical liquid situation similar to a single edge of a quantum spin Hall insulator: spins-up are only allowed to travel right, while spins-down only travel left [95]. Coupling a conventional *s*-wave superconductor by proximity to the semiconductor nanowire adds particle-hole symmetry and produces Majorana fermion bound states at the ends of the nanowire [96], [97]. Majorana fermions should manifest themselves as peaks in conductance at zero bias, which were indeed observed in a tunneling experiment (Fig. 4(b)). Interestingly, when the external magnetic field is aligned with the internal Rashba field, no subband hybridization occurs and the gap at zero momentum does not open. In this case Majorana fermions are not expected and the zero-bias peak vanishes [98].

Another recent experiment has attempted to look for Majorana fermions in chains of magnetic atoms on a superconductor surface [99]. The ingredients of this approach are essentially the same as with semiconductor nanowires, but with several remarkable differences. First, the SO interaction in this case is proposed to originate from the superconductor (Pb used in the experiment has a strong intrinsic SO interaction). Second, time-reversal symmetry is broken by the magnetization of the chain rather than by an external field. Experimentally, zero-bias states were detected by an STM experiment at the ends of the atomic chains (Fig. 4(c)). A similar outcome would have been possible if the magnetic atoms spontaneously formed a spin helix thereby creating a synthetic SO interaction, like in the case of cold atoms discussed below, along the chain [100].

The significance of this research direction goes beyond the discovery of new topological classes of matter. In the case of helical edge states, the absence of backscattering at zero magnetic field may in the future play a role in reducing dissipation in spintronic and electronic circuits. Majorana bound states are expected to exhibit non-Abelian exchange statistics [101]. This means that, as opposed to conventional fermions and bosons, when one Majorana bound state is moved around another following a closed loop, the system undergoes a transition to a new ground state of distinctly different charge. This non-Abelian property is yet to be demonstrated experimentally. If realized, it may open the door to the realization of topological quantum computing, in which error-protected quantum operations are performed by moving Majorana fermions around each other, also known as braiding [102].

Topological states that do not involve SO coupling have been proposed theoretically [103]. Those states rely on other crystalline symmetries that play the role of SO interaction. However, SO interaction has clearly been essential for the emergence of the first several waves of experiments and concepts in the field of topological condensed matter systems.

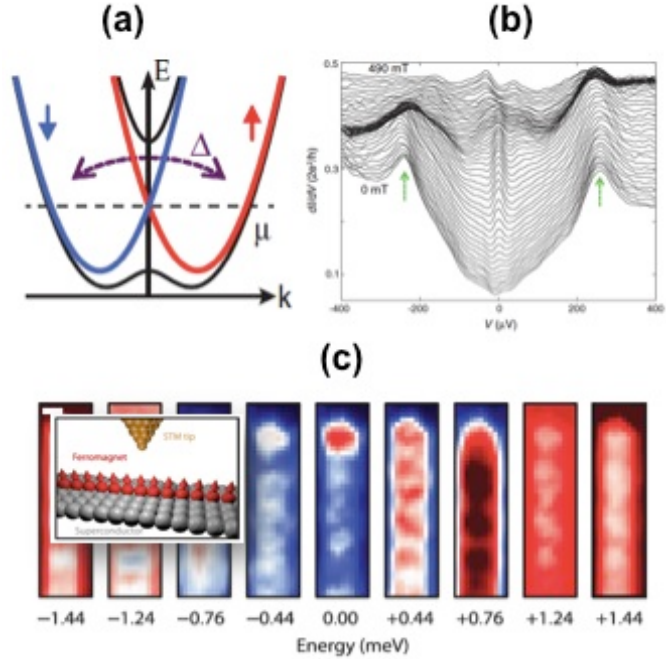

**Figure 4| Majorana fermions:** (a) Schematics of the band structure arising from the interplay between topological Dirac cone and superconductivity; (b) Magnetic field-dependent spectroscopy of the device displaying the induced gap peaks (green arrow) (from Ref. [98]) (c) STM imaging demonstrating the emergence of Majorana fermion in a ferromagnetic-superconductor hybrid structure shown in inset (from Ref. [99]).

## Low dimensional Dirac systems

Another class of materials in which topological phases have been identified are the so-called honeycomb crystals, which present striking similarities with two-dimensional relativistic massive particles. The realization of systems displaying Dirac-type hamiltonians dates back to the exploration of superfluidity in $^3$He [104], [105] and $d$-wave superconductivity [106] (see also Ref. [107]). In such systems, the momentum of the quasiparticle is directly coupled to their Nambu pseudospin, i.e., a spinor formed by electron and hole part (see Box 3), which results in an analog to the quantum Hall effect in the absence of magnetic fields [104]. More recently, the realization of two-dimensional honeycomb crystals, displaying a Dirac cone at the two valleys at the high symmetry K and K' points, has introduced a new paradigm for effective relativistic condensed matter physics (see Fig. 3 (a,b)).

The physics of Dirac particles introduces a wealth of thought-provoking phenomena among which Klein tunneling and Zitterbewegung are probably the more illustrative. Klein tunneling is the absence of backscattering from a potential (defects and impurities) due to the penetration of negative energy particles into the barrier, resulting in large mobilities [108], [109] as mentioned above in the context of topological insulators. Zitterbewegung is literally the jittering or trembling motion of the carrier, which is a direct consequence of the locking between the momentum $\vec{p}$ and the (pseudo)spin momentum. From the perspective of Dirac physics, this effect stems from time-dependent interferences between positive and negative energy particles (electrons and positrons).

Although the existence of Dirac cones in graphene was realized about 60 years ago [110], it has only fully reached its potential recently with the by now famous rise of experimental graphene [111]. In this system, the pseudospin corresponds to the two lattice sites composing the motif of the crystal. Formally, the (pseudo) SO coupling does not arise from structural inversion asymmetry and hence cannot be referred to as Rashba SO coupling *per se*. In fact, one of the virtues of the development of graphene has been to shed light on a variety of novel solid-state materials displaying a Dirac cone at low energy (see Box 3). These materials are two-dimensional graphene-like crystals, such as silicene, germanene, stanene, *h*-boron-nitride or transition metal-based dichalcogenides (TMDCs) such as $MoS_2$, $WSe_2$ etc. They all display a Dirac cone at their high symmetry K and K' points. The pseudospin involved in the Dirac cone can be a superposition of lattice sites (two-dimensional diamonds) or a superposition of unperturbed orbitals (TMDCs). They present a unique playground to explore the Dirac world at the solid state level as they possess very flexible properties.

In analogy with the topological insulators presented in the previous section, honeycomb crystals also display a quantum Hall effect in the insulating regime. The subtlety is that since the number of Dirac cones is even in this case, the nature of the quantum Hall effect depends on the nature of the gap. For instance, if the symmetry between the two sublattices is broken (as in the case of TMDCs or *h*-boron-nitride), the two valleys contribute to an opposite quantum Hall conductivity which results in a quantum *valley* Hall effect [112] (see Fig. 3(d)), i.e. a charge neutral current (somewhat similar to the quantum spin Hall effect). Similarly, the gap induced by SO coupling results in a quantum spin Hall effect, but a vanishing quantum valley Hall effect [113] (as in silicene or germanene). Interestingly, the quantum spin Hall effect is accompanied by spin-polarized edge currents similar to the ones observed in topological insulators [87], [114]. Finally, a last interesting situation is obtained when coupling the honeycomb crystal to an antiferromagnetic insulator [115] (i.e. both spatial and time-reversal symmetries are broken), thereby realizing the original Haldane model [116] (see Fig. 3(c)). In this case, the contribution of the two valleys to the quantum Hall effect do not compensate each other anymore, resulting in a quantum anomalous Hall effect, i.e. a quantized transverse charge current.

The observation of quantum spin, valley or anomalous Hall effects in low dimensional Dirac systems (i.e. not topological insulators) is still very challenging and to the best of our knowledge, no topological quantum Hall effect at zero magnetic field has been reported in these materials. Nonetheless, the observation of a large spin Hall effect ($\theta_H \approx 20\%$) in graphene attributed to extrinsic SO coupling [117], [118], as well as the realization of a magnetic field-induced quantum spin Hall effect constitute promising progress towards SO coupled transport. In this latter experiment, in-plane and perpendicular magnetic fields are combined to generate spin-polarized charge neutral edge currents [119]. The existence of large intrinsic Rashba SO coupling in silicene [120], germanene [121] and possibly stanene [122], as well as the recent demonstration of large extrinsic Rashba SO coupling (~20meV) in graphene [123], [124] could be the premises to a breakthrough in this field since Rashba SO coupling enables the coupling between pseudospin, spin and momentum degrees of freedom.

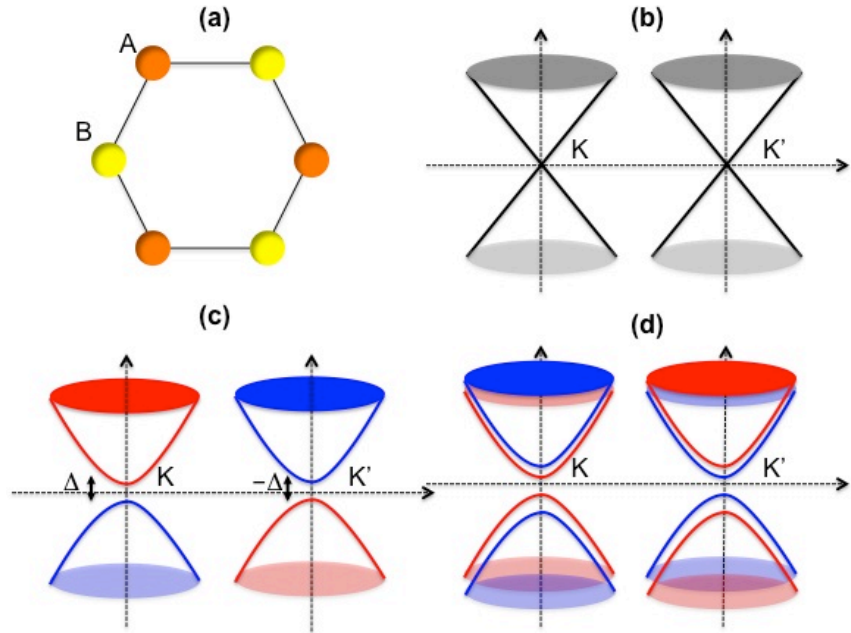

**Figure 5| Low dimensional Dirac materials:** (a) Schematics of the hexagonal honeycomb crystal. Sites A and B constitute the pseudospin states. (b) Low energy band structure of graphene, constituted of two Dirac cones located at the K and K' points of the Brillouin zone. The chirality of the bands are opposite at these two points. (c) Low energy band structure of Haldane model: the gap is $+\Delta$ at K and $-\Delta$ at K' points, enabling the emergence of quantum anomalous Hall effect, i.e. a net flow of charges. (d) Low energy band structure of a transition metal dichalcogenide monolayer. The inversion asymmetry induces a gap and SO coupling induces a supplementary spin splitting.

Meanwhile, intense efforts have been achieved towards the manipulation of valley polarization and the realization of the valley Hall effect, i.e. the generation of a transverse charge neutral current induced by the Berry curvature, which could be used as a new functional degree of freedom [125], [126]. Breaking spatial inversion symmetry by hybridizing graphene with *h*-boron-nitride substrate [127] has enabled the observation of charge neutral current in graphene [128], [129], possibly associated with spin or valley Hall effect.

The case of TMDCs is worth special attention [130], [131], [132]. Their large band gap (~1.5-2eV) enables optical control of the valley population using resonant light [112], as demonstrated experimentally [133]–[135]. Furthermore, due to their strong SO coupling, valley and spin are coupled (this is particularly true in WSe$_2$ for instance). The light-induced valley polarization can be used to generate valley Hall effect [136] or spin-valley coupled photogalvanic effect [137]. Inversely, electrically-driven emission of circularly polarized light has been demonstrated recently in a *p-n* junction geometry in WSe$_2$ monolayer [138]. The influence of symmetry breaking in TMDC bilayers is also currently attracting much attention. The gap at K and K' valleys can be modified by applying an inversion symmetry breaking perpendicular electric field, as optically probed in bilayer $MoS_2$ [139].

**Box 3| The Concept of Pseudospin**

**Pseudospin in materials**

While the spin angular momentum is the quantized intrinsic angular momentum of a particle, many other physical quantities act as an effective spin ½ system dubbed *pseudospin*. This concept was originally introduced by Heisenberg to describe the structure of the atomic nucleus as composed of neutrons and protons, modeled as two states of the same particle [140]. In this context, a pseudospin is a coherent superposition of two quantum states and is described in terms of Paul matrices for spin ½, $\vec{\sigma} = (\sigma_x, \sigma_y, \sigma_z)$ [141]. While Nambu pseudospin has been introduced decades ago to describe quasiparticles in superconductors, recent developments in the physics of SO coupled transport have identified new degrees of freedom that can be accounted for within the pseudospin language [107]. In hexagonal two-dimensional diamonds the pseudospin is composed of the sublattices [110] while in TMDC, it describes the valence and conduction bands of the transition metal [112]. In Van der Waals bilayers, when the layer index is a good quantum number, a layer pseudospin can also be identified [142]. Finally, in cold-atom systems, the spin ½ pseudospin is defined by two hyperfine split states that can be, e.g., coherently coupled by a laser [143].

The concept of pseudospin is useful in predicting and interpreting transport properties of the various systems mentioned above, in particular when a Rashba-type pseudospin-orbit coupling is present [107]. Nonetheless, it has an important limitation: the nature of the pseudospin (sublattice, layer index etc.) is a material property, not an intrinsic property of the carrier like the spin degree of freedom (except in the case of cold atoms). Therefore, it may not be continuous at the interfaces between different materials [142].

**Pseudospin-Orbit Coupling: The example of cold atoms**

To exemplify the use of the pseudospin concept, let us consider a simple example of cold atoms. We imagine that two atomic hyperfine states – labeled $\Psi_\uparrow$ and $\Psi_\downarrow$, and selected from a large integer-spin hyperfine multiplet – representing the two states of the pseudospin, are coupled by Raman lasers. The single-particle Hamiltonian is then given by

$$\hat{H} = -\frac{\hbar^2}{2m}\left(\frac{\partial^2}{\partial x^2} + \frac{\partial^2}{\partial y^2}\right)\begin{pmatrix} 1 & 0 \\ 0 & 1 \end{pmatrix} - \frac{\Delta}{2}\begin{pmatrix} 1 & 0 \\ 0 & -1 \end{pmatrix} - \frac{\Omega}{2}\begin{pmatrix} 0 & e^{ik_p x} \\ e^{-ik_p x} & 0 \end{pmatrix}, \quad \text{(B.2)}$$

where $\Omega$ is determined by the strength of the lasers (that are chosen to point along the *x* direction), and $k_p$ is the photon wave vector. The Zeeman splitting $\Delta$ is controlled by the external magnetic field, and *m* is the mass of the atoms. We have ignored motion in the *z* direction, which is justifiable as long as the confining potential is tight in this direction.

> The Schrödinger equation for the atoms, $\hat{H}\Psi = E\Psi$, with $\Psi = (\Psi_\uparrow(x,y), \Psi_\downarrow(x,y))$, is now rotated according to $\Psi = Q\phi$. Here $Q$ is the 2x2 matrix that diagonalizes the spin part of the Hamiltonian such that
>
> $$Q^{-1}\left[-\frac{\Delta}{2}\begin{pmatrix}1 & 0 \\ 0 & -1\end{pmatrix} - \frac{\Omega}{2}\begin{pmatrix}0 & e^{ik_p x} \\ e^{-ik_p x} & 0\end{pmatrix}\right]Q = -\frac{1}{2}\sqrt{\Omega^2 + \Delta^2}. \quad (B.3)$$
>
> The kinetic-energy part of the Schrödinger equation transforms according to
>
> $$-\frac{\hbar^2}{2m}Q^{-1}\left(\frac{\partial^2}{\partial x^2} + \frac{\partial^2}{\partial y^2}\right)\begin{pmatrix}1 & 0 \\ 0 & 1\end{pmatrix}(Q\phi). \quad (B.4)$$
>
> Due to the laser field we have that $Q=Q(x,y)$, leading to terms linear in momentum in the Schrödinger equation for $\phi$. One of these terms is a pseudo-SO coupling $\gamma\sigma_x \partial/\partial x \propto \gamma\sigma_x p_x$, with $\gamma = \hbar k_p \Omega/2m\sqrt{\Omega^2 + \Delta^2}$. The Rashba pseudo-SO coupling can be engineered using more involved laser-coupling schemes [144].

**Rashba physics with cold-atom systems**

An emerging direction for exploring the physics of SO interactions is the development of cold-atom systems. These are ultracold (0.1-10 $\mu$K) clouds of typically up to $10^9$ neutral alkali atoms that are trapped in a magnetic or optical confining potential, using the Zeeman or AC Stark effect, respectively. Due to this method of trapping, the atoms are essentially isolated from their environment and do not experience disorder or lattice vibrations, contrary to electrons in solids. The atoms could be either fermions or bosons and interact via short-range interactions as opposed to the long-ranged Coulomb interactions felt by electrons. Various properties of cold-atom systems, such as confining potential, temperature, density and strength of interactions, can be experimentally tuned. Successes that have been achieved with cold-atom systems are in large part due to this amount of tunability and the fact that these systems explore new physical regimes as compared to electrons in solid-state materials or other condensed-matter systems. Examples of groundbreaking experiments in this respect are the exploration of the crossover between Bose-Einstein and Bardeen-Cooper-Schrieffer regimes of fermion pair condensations [145], the observation of vortex unbinding in two-dimensions [146], and the Mott insulator-to-superfluid quantum phase transition [147].

For a gas of alkali-metal atoms, the role of spin is played by the hyperfine spin degrees of freedom of the atoms (see Box 3). SO coupling here refers to coupling between the motion of the entire neutral atom to its hyperfine spin, and not to the coupling between the orbital momentum of the valence electron of the atom to its spin. Because the atoms are neutral this SO coupling arises in a different manner than for electrons and has to be engineered. It is therefore referred to as *synthetic* SO coupling (see Box 3).

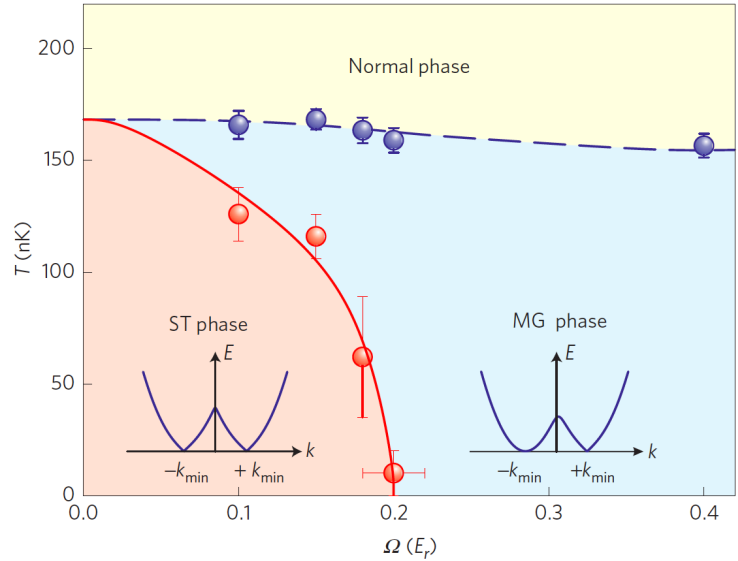

**Figure 6| Finite-temperature phase diagram of a SO coupled Bose gas.** The lines separate stripe phases (ST), magnetized phases (MG) and a normal phase. The vertical axis corresponds to temperature and the horizontal axis corresponds to the strength of the lasers (in units of the recoil energy) that determine the strength of the SO coupling. Taken from Ref. [148].

Very recent experimental efforts have succeeded in creating this synthetic SO coupling. The first experiment concerned bosons and a transition between phase-mixed and phase-separated dressed spin states was observed in the Bose-Einstein-condensed regime [149]. Subsequent experiments on coupling the linear motion of bosonic atoms to their hyperfine spin succeeded in demonstrating strong synthetic orbital magnetic fields [150], changes in the dipole collective mode due to SO coupling [151], Zitterbewegung [152], and, very recently, in mapping out the finite-temperature phase diagram (see Fig. 6) [148]. Similar efforts with fermionic atoms [153] demonstrated the emergence of a SO gap in these systems (see Fig. 7) [154], as well as the realization of the Haldane model [155].

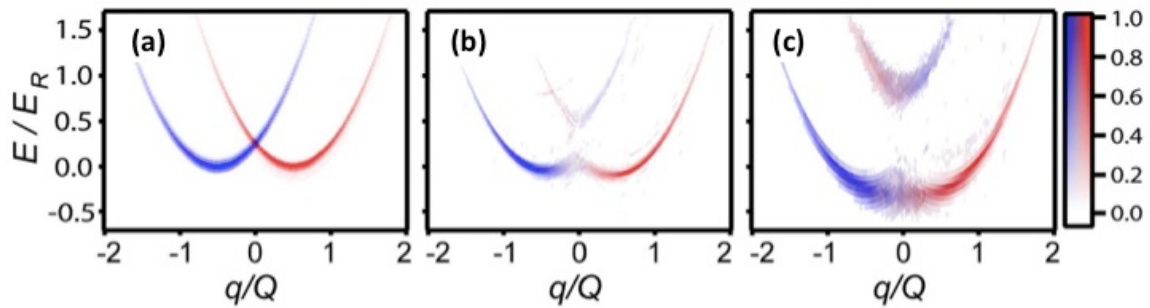

**Figure 7: Results of measurements of the SO coupled dispersion and SO gap in a gas of fermionic atoms by spin-injection spectroscopy.** In these measurements, a radio-frequency pulse transfers atoms from the SO coupled states to an unoccupied hyperfine state. Measuring the number of atoms in this initially unoccupied state as a function of the energy of the pulse gives information on the energy dispersion of the trapped atoms (or, more generally, their spectral function, similar to ARPES in electronic systems). Here, the energies are in units of the recoil energy of the laser that engineers the synthetic SO

coupling, and the momentum $q$ is in units of the photon momentum $Q$ ($k_p$ in the example of Eq. (B.2)). Taken from Ref. [154].

Ongoing and future efforts are dedicated to engineering more complicated synthetic SO coupling schemes. This paves the way for a variety of fundamental research. First, fermionic systems may serve as controlled quantum simulators for electrons and allow for singling out particular effects, such as the competition between SO coupling and many-body interactions. Second, SO coupling in cold-atom systems gives rise to new phenomena arising from coupling between hyperfine spin and linear motion. One example of this is the observation of synthetic partial waves in interatomic collisions [156]. Finally, both fermionic and bosonic systems enable engineering completely novel states of matter that have no analogue in the solid state. Examples of these are SO-coupled Mott insulators and superfluids that arise in systems of strongly-interacting SO coupled cold atoms in optical lattices [157]–[161]. Most spectacular, perhaps, is the outlook of SO-coupled cold atoms to realize a host of exotic phases known as bosonic topological insulators [162], [127]. These phases are reminiscent of electronic topological insulators in that they support edge states, but contrary to electronic topological states, arise only in the presence of interactions.

**Summary and Outlook**

The advancement of research in SO coupled transport of inversion asymmetric systems has been extremely creative in the past ten years. Wide areas of physics and materials science, traditionally treated on different footing (metallic spintronics, Van der Waals materials, cold atom systems), are now converging under the umbrella of spin-orbitronics. Traditional spintronics has already been through two major revolutions in its history (giant magnetoresistance and spin transfer torques) and is currently experiencing its third one thanks to the development of chiral magnetic structures. On the low dimensional side, while it is not entirely clear whether graphene will eventually keep its promises, novel low dimensional systems such as TMDC, silicene, germanene, stanene and topological insulators are offering even broader opportunities for materials design. Finally, Rashba-like SO coupling empowers unique topological properties that are, for example, expected in superconductors and cold atom systems. Exotic states of matter such as bosonic topological states stabilized by interactions will surely keep the heat on for the next decades.

While we chose to focus this review on a selected numbers of topics whose development is highly promising, additional subjects deserve attention but could not be included. The electrical and optical control of spin in semiconductors is a vast area of which we could only give an imperfect account [164]. For instance, electron dipole spin resonance could have been the subject of a much deeper presentation. We also wish to mention that novel materials displaying extremely large Rashba-type SO coupling in their bulk are currently drawing major interest (such as BiTeI polar semiconductors or $R_2Ir_2O_7$ pyrochlores [165], [166]), paving the way towards to experimental realization of Weyl semimetals and other exotic phases [167]. Finally, concepts related to Rashba SO coupling in electronic systems have also been recently extended to optical properties of chiral biological systems [168].

## Acknowledgement

The authors acknowledge useful discussions with E.I. Rashba, Lars Fritz, Di Xiao and A.H. MacDonald. A.M. was supported by the King Abdullah University of Science and Technology (KAUST). H.C.K was supported by the KIST and KU-KIST Institutional Program. J.N. acknowledges support by the Grants-in-Aid from the Japan Society for the Promotion of Science (JSPS; no. 22226001). S.F. acknowledges ONR BRC on Majorana Fermions, NSF, Sloan Foundation, Charles E. Kaufman foundation, Nanoscience Foundation. R. D. is supported by the Stichting voor Fundamenteel Onderzoek der Materie (FOM), the European Research Council (ERC) and is part of the D-ITP consortium, a program of the Netherlands Organisation for Scientific Research (NWO) that is funded by the Dutch Ministry of Education, Culture and Science (OCW).